\documentclass[aps,pra,showpacs,groupedaddress,twocolumn]{revtex4}

\usepackage{graphicx}
\usepackage{dcolumn}
\usepackage{bm}
\usepackage[usenames,dvipsnames]{color}
\usepackage{amsmath}
\usepackage[hypertex]{hyperref}
\usepackage{amsfonts}
\usepackage{amsthm}
\usepackage{amssymb}
\usepackage{mathrsfs}
\usepackage{subfigure}
\usepackage{ulem}

\begin{document}

\title{Quantum discord in quantum random access codes and its connection with dimension witness}

\author{Yao Yao}

\author{Hong-Wei Li}

\author{Xu-Bo Zou}
\email{xbz@ustc.edu.cn}

\author{Jing-Zheng Huang}

\author{Chun-Mei Zhang}

\author{Zhen-Qiang Yin}

\author{Wei Chen}

\author{Guang-Can Guo}

\author{Zheng-Fu Han}
\email{zfhan@ustc.edu.cn}
\affiliation{Key Laboratory of Quantum Information,University of Science and
Technology of China,Hefei 230026,China}

\date{\today}

\begin{abstract}
We exploit quantum discord (and geometric discord) to detect quantum correlations present in a well-known communication model called
quantum random access codes (QRACs), which has a variety of applications in quantum information theory. In spite of the fact that there is no entanglement between the two parts involved in this model, analytical derivation shows that the quantum discord is nonzero and highlights that quantum discord might be regarded as a figure of merit to characterize the quantum feature of QRACs, since this model has no classical counterparts. To gain further insight, we also investigate the dynamical behavior of quantum discord under some specific state rotations.
In two-state case, the connection between quantum discord and dimension witness is graphically discussed and
intriguingly our results illustrate that these two quantities are monotonically related to each other.
For state encodings in real $|0\rangle-|1\rangle$ plane, we derive an explicit analytical expression of the geometric discord
and find that geometric discord reaches the maximal value for the optimal encoding strategy.
However, for arbitrary state encodings in Bloch sphere, our numerical simulations reveal that
maximal geometric discord could not coincide with optimal $2\rightarrow1$ QRAC.
\end{abstract}


\maketitle
\section{INTRODUCTION}
Since the advent of the concept of quantum discord \cite{Ollivier2001,Henderson2001}, a great deal of endeavor \cite{Dakic2010,Luo2010a,Luo2008a,Modi2010,Oppenheim2002}
has been devoted to classifying and quantifying the quantum correlations which do not necessarily involve quantum entanglement. It is now well-known
that almost all quantum states possess nonclassical correlations \cite{Ferraro2010}. Therefore, the significance of quantum discord beyond entanglement
partly lies in the fact that it can be utilized as an informational-theoretical tool to analyze the quantum correlations contained in separable states,
since in these circumstances entanglement can by no means be regarded as the physical resource for the realization of certain quantum information tasks.
Along this line of thought, A. Datta \textit{et al.} drew the community's attention to the deterministic quantum computation with one quantum bit, or
the so called DQC1 model, in which the quantum discord other than entanglement is suggested to be the figure of merit for characterizing the resources
present in this computational model \cite{Datta2008,Lanyon2008}. Recently, it has been reported that quantum correlations (quantified by some discord-like
measures) also play a vital role in some other quantum tasks such as remote state preparation \cite{Dakic2012,Tufarelli2012} and entanglement distribution
using separable states \cite{Streltsov2012a,Chuan2012}.

In particular, another quantum task demanding for quantum correlations but not entanglement is quantum key distribution (QKD) \cite{Bennett1984}.
This motivates us to investigate other quantum communication models which are not based on entanglement, while what comes into
our sight is quantum random access code protocol \cite{Wiesner1983,Ambainis1999,Ambainis2002,Hayashi2006}. QRACs have a variety of applications in areas
ranging form quantum communication complexity \cite{Klauck2001}, network coding \cite{Hayashi2007}, information causality \cite{Pawlowski2009},
to security proof of QKD protocol \cite{Pawlowski2011}. Following the spirit of $n\rightarrow1$ quantum random access codes,
we have proposed a semi-device-independent random-number expansion protocol in our previous work \cite{Li2011,Li2012}. In this protocol
no entanglement is required and the randomness can be guaranteed only by the two-dimensional quantum witness violation, which is in
sharp contrast to the random-number-generation protocol certified by the Bell inequality violation \cite{Pironio2010}.
\begin{figure}[htbp]
\begin{center}
\includegraphics[width=.40\textwidth]{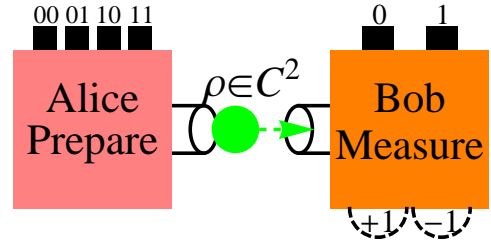} {}
\end{center}
\caption{(Color online) The sketch of $2\rightarrow1$ quantum random access code. Alice encodes her randomly chosen 2 classical bits $a\in\{00,01,10,11\}$
into 1 qubit $\rho_a$ and sends it to Bob. To decode the required bit, Bob performs some measurement on the received qubit depending on his input bit $y\in\{0,1\}$
with the measurement results denoted as $b\in\{+1,-1\}$ (which in the computational basis can be represented by $b\in\{0,1\}$).
}
\label{2-1}
\end{figure}

In this work, we focus on the $2\rightarrow1$ QRAC, which is sketchily depicted in Figure \ref{2-1} (a more detailed description will be given
in next section). Since there exists no entanglement in this model and no classical counterpart, we exploit quantum discord (and geometric measure
of discord) to characterize the nonclassical nature of QRACs. Indeed, our analytical results show that the quantum discord is nonzero, and more
fascinatingly, reaches the maximal value for the optimal encoding for $2\rightarrow1$ QRAC. To go deeper into the state encodings, we also
step forward to study the dynamical behavior of quantum discord regarding some possible state variations (rotations, in fact). Except for
the above-mentioned intrinsic interest of QRACs, we also try to clarify the relationship between the two-dimensional quantum witness
\cite{Pawlowski2011,Gallego2010,Brunner2008} and quantum correlation.
By numerical evaluations, it has turned out that these two quantities are monotonically related to each other, which may indicate
the randomness associated with the dimension witness may originate from nonclassical correlations.

The outline of this paper is as follows. In Sec. II, we briefly review the notations and definitions used throughout this paper.
In Sec. III, we turn to analyze the quantum correlations in $2\rightarrow1$ QRAC, including the original quantum discord and
the geometric version. In Sec. IV, we go further to investigate the dynamical behavior of quantum discord with respect to
state rotations. Sec. V is devoted to the discussion and conclusion.

\begin{figure}[htbp]
\begin{center}
\includegraphics[width=.40\textwidth]{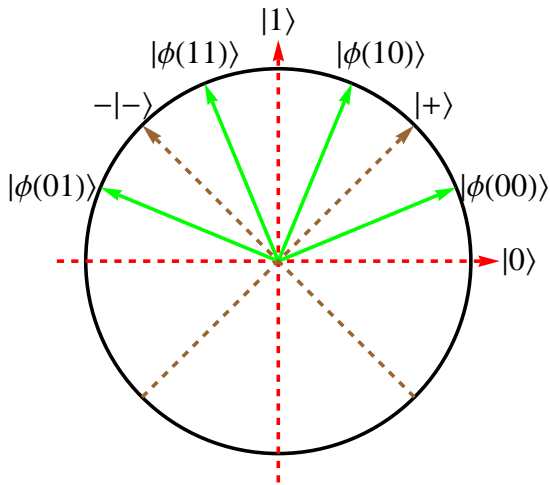} {}
\end{center}
\caption{(Color online) Optimal $2\rightarrow1$ quantum random access coding in $|0\rangle-|1\rangle$ plane representation \cite{Hayashi2006}.
}
\label{Bloch}
\end{figure}

\section{NOTATIONS AND DEFINITIONS}
\textit{\textbf{Quantum random access codes.}}
The idea behind QRACs was first raised by Stephen Wiesner \cite{Wiesner1983} in 1983 and was labeled as \textit{conjugate coding} at that time.
More than a decade later, these codes were re-discovered by Ambainis \textit{et al.} in Ref \cite{Ambainis1999,Ambainis2002} and represented
in the standard form: $(n,m,p)$-QRA codings. Here, the notion of $(n,m,p)$-QRACs is adopted to denote the task in which the sender (Alice)
encodes $n$ classical bits into $m$-qubit states in order that the receiver (Bob) can recover any one bit of the initial n bits with probability
at least $p$. To exhibit the advantage of quantum codings over classical encodings, Ambainis \textit{et al.} presented an exact example for
$\big(2, 1, \frac{1}{2}(1+\frac{1}{\sqrt{2}})\big)$-QRAC and referred to its straightforward generalization to $\big(3, 1, \frac{1}{2}(1+\frac{1}{\sqrt{3}})\big)$-QRAC
by Chuang \cite{Ambainis1999}, which are just the optimal codings for cases $n=2$ and 3. However, Hayashi \textit{et al.} proved there is no
$(4, 1, p)$-QRAC such that $p$ is strictly greater than 1/2 \cite{Hayashi2006} (For history and applications about QRACs, we refer the readers
to an extended work \cite{Ambainis2008}).

From now on, we concentrate on $2\rightarrow1$ QRAC. To begin with, let us introduce the $\big(2, 1, \frac{1}{2}(1+\frac{1}{\sqrt{2}})\big)$-QRA coding
strategy. Alice encodes her two random bits $a_1a_2\in\{0,1\}^2$ into one qubit $\rho_{a_1a_2}=|\phi(a_1a_2)\rangle\langle\phi(a_1a_2)|$ where
\begin{align}
\label{encoding}
|\phi(00)\rangle &= \cos(\frac{\pi}{8})|0\rangle+\sin(\frac{\pi}{8})|1\rangle,\nonumber\\
|\phi(01)\rangle &= \cos(\frac{7\pi}{8})|0\rangle+\sin(\frac{7\pi}{8})|1\rangle,\nonumber\\
|\phi(10)\rangle &= \cos(\frac{3\pi}{8})|0\rangle+\sin(\frac{3\pi}{8})|1\rangle,\nonumber\\
|\phi(11)\rangle &= \cos(\frac{5\pi}{8})|0\rangle+\sin(\frac{5\pi}{8})|1\rangle,
\end{align}
To extract the required bit, Bob performs the two projective measurements as follows
\begin{align}
\label{decoding}
M_1 &= \{M_0^0=|0\rangle\langle0|,\, M_0^1=|1\rangle\langle1|\},\nonumber\\
M_2 &= \{M_1^0=|+\rangle\langle+|,\, M_1^1=|-\rangle\langle-|\},
\end{align}
with $|\pm\rangle=\frac{1}{\sqrt{2}}(|0\rangle\pm|1\rangle)$. In Figure \ref{Bloch}, we explicitly illustrate the optimal state encodings and
decodings in $|0\rangle-|1\rangle$ plane representation. It is easy to see that the probability that Bob successfully recovers any of Alice's two bits is
$\cos(\frac{\pi}{8})^2=\frac{1}{2}(1+\frac{1}{\sqrt{2}})\approx0.85$. On the contrary, there exists no $2\rightarrow1$ classical encoding for any
$p>\frac{1}{2}$ \cite{Ambainis1999}. The gap between quantum and classical encodings motivates us to investigate the quantum correlations
in quantum encodings, which is very likely responsible for the advantage over classical encodings.
In fact, Alice's coding strategy can be written as a mixture of
four product states
\begin{align}
\label{state}
\rho_{AB}=&\frac{1}{4}(|00\rangle\langle00|\otimes\rho_{00}+|01\rangle\langle01|\otimes\rho_{01}\nonumber\\
&+|10\rangle\langle10|\otimes\rho_{10}+|11\rangle\langle11|\otimes\rho_{11}).
\end{align}
Obviously, there exists no entanglement between the natural bipartite split, and this $4\otimes2$ \textit{classical-quantum} state
is just our starting point for the later analysis.

\textit{\textbf{Quantum discord.}}
The quantum discord is proposed by Ollivier and Zurek as an informational-theoretical measure of the quantumness of correlations,
which originates from the inequivalence of two classically identical expressions of the mutual information in the quantum realm \cite{Ollivier2001}.
For a given composite system $\rho_{AB}$
\begin{align}
\mathcal{D}_A(\rho):=\mathcal{I}(\rho)-\mathcal{J}(\rho|\{\Pi_A^k\}),
\end{align}
where $\mathcal{I}(\rho)=S(\rho_A)+S(\rho_B)-S(\rho)$ denotes the quantum mutual information, $S(\rho)=-Tr(\rho\log_2\rho)$
is the von Neumann entropy, $\rho_{A(B)}=Tr_{B(A)}(\rho)$ represent the reduced states for subsystem A(B) and $\mathcal{J}(\rho|\{\Pi_A^k\})$
is suggested by Henderson and Vedral as a measure to quantify the classical correlation \cite{Henderson2001}
\begin{align}
\mathcal{J}(\rho|\{\Pi_A^k\}):&= S(\rho_B)-\min_{\{\Pi_A^k\}}S(\rho|\{\Pi_A^k\}),\nonumber\\
&=S(\rho_B)-\min_{\{\Pi_A^k\}}p_kS(\rho_B^k),
\end{align}
where $p_k=Tr(\Pi_{A}^{k}\rho)$ and $\rho_B^k=Tr_A(\Pi_{A}^{k}\rho)/p_k$, and the minimum is taken over all von Neumann measurements
$\{\Pi_A^k\}$ to eliminate the dependence on specific measurement. Although
much endeavor has been devoted to calculating quantum discord for the two-qubit states \cite{Luo2008b,Ali2010,Girolami2011,Chen2011},
analytical results for high-dimensional systems are rarely to be found in the literature \cite{Datta2008,Chitambar2011}.
To compute the quantum discord of state (\ref{state}), we should resort to the original formula of discord defined here.

\textit{\textbf{Geometric discord.}}
Based on the Hilbert-Schmidt norm, Daki\'{c} \textit{et al.} introduced the following geometric measure of quantum discord \cite{Dakic2010}
\begin{align}
\mathcal{D}^{G}_A(\rho):=\min_{\chi\in\Omega}\|\rho-\chi\|^2,
\end{align}
where $\Omega$ denotes the set of zero-discord states and $\|\rho-\chi\|^2=Tr(\rho-\chi)^2$ is the square of
Hilbert-Schmidt norm. For the two-qubit case, an analytic form of geometric discord can be obtained
\begin{align}
\mathcal{D}^{G}_A(\rho)=\frac{1}{4}(\|\vec{x}\|^2+\|T\|^2-\lambda_{max}),
\end{align}
where $x_i=Tr(\sigma^A_i\rho)$ are components of the local Bloch vector for subsystem A, $T_{ij}=Tr(\sigma^A_i\sigma^B_j\rho)$
are components of the correlation matrix, and $\vec{x}:=(x_1,x_2,x_3)^t$, $T:=(T_{ij})$, $\lambda_{max}$ is the largest eigenvalue of
the matrix $K=\vec{x}\vec{x}^t+TT^t$ (here the superscript t denotes transpose).

It is worth mentioning that, Luo and Fu presented a simplified version of the geometric discord \cite{Luo2010a}
\begin{align}
\mathcal{D}^{G}_A(\rho)=\min_{\Pi_A}||\rho-\Pi_A(\rho)||^2,
\end{align}
where the minimum is over all von Neumann measurements $\Pi_A=\{\Pi_A^k\}$ on subsystem A. Following the treatment method
in \cite{Luo2010a}, the authors of Ref. \cite{Rana2012} and \cite{Hassan2012} derived a tight lower bound to
the geometric discord of arbitrary $m\otimes n$ states
\begin{align}
\label{bound}
\mathcal{D}^{G}_A(\rho)\geq\frac{2}{m^2n}(\|\vec{x}\|^2+\frac{2}{n}\|T\|^2-\sum_{i=1}^{m-1}\lambda_i^\downarrow),
\end{align}
where $\lambda_i^\downarrow$ are the eigenvalues of $G=\vec{x}\vec{x}^t+\frac{2TT^t}{n}$ listed in decreasing order (counting multiplicity)
and here $\vec{x}=(x_1,x_2,...,x_m)^t$, $T=(T_{ij})$ are given by
\begin{align}
\label{decomposition}
x_i&=\frac{m}{2}Tr(\rho\tilde{\lambda_i}\otimes I_n)=\frac{m}{2}Tr(\rho_A\tilde{\lambda_i}),\nonumber\\
T_{ij}&=\frac{mn}{4}Tr(\rho\tilde{\lambda_i}\otimes \tilde{\lambda_j}).
\end{align}
with $\tilde{\lambda_i},\tilde{\lambda_j}$ being the generators of $SU(d)$ for corresponding dimension $d=m,n$ \cite{generator1}.
It is remarkable that the lower bound in Eq. (\ref{bound}) is saturated by all $2\otimes n$ states (with the measurement on the qubit)
\cite{Rana2012,Hassan2012} (the same result was also obtained in Ref. \cite{Vinjanampathy2012}). This analytical formula
can be directly applied to our case.

\section{CORRELATION ANALYSIS IN QRAC}
Equipped with these concepts and formulas, we are now in the position to analyze the quantum correlation in $2\rightarrow1$ QRAC,
measured by quantum discord (QD) and geometric measure of discord (GD) respectively.

\subsection{Quantum discord}
First, we notice that state (\ref{state}) (an $8\times8$ matrix in fact) can be cast into a block diagonal matrix
\begin{equation}
\rho_{AB}=
\left(\begin{array}{cccc}
\rho_{00} & 0 & 0 & 0 \\
0 & \rho_{01} & 0 & 0 \\
0 & 0 & \rho_{10} & 0 \\
0 & 0 & 0 & \rho_{11}
\end{array}\right),
\end{equation}
where
\begin{align}
\label{double}
\rho_{00}&=\frac{1}{2}
\left(\begin{array}{cc}
1+\cos\frac{\pi}{4} & \sin\frac{\pi}{4}  \\
\sin\frac{\pi}{4} & 1-\cos\frac{\pi}{4}
\end{array}\right),\nonumber\\
\rho_{01}&=\frac{1}{2}
\left(\begin{array}{cc}
1+\cos\frac{7\pi}{4} & \sin\frac{7\pi}{4}  \\
\sin\frac{7\pi}{4} & 1-\cos\frac{7\pi}{4}
\end{array}\right),\nonumber\\
\rho_{10}&=\frac{1}{2}
\left(\begin{array}{cc}
1+\cos\frac{3\pi}{4} & \sin\frac{3\pi}{4}  \\
\sin\frac{3\pi}{4} & 1-\cos\frac{3\pi}{4}
\end{array}\right),\nonumber\\
\rho_{11}&=\frac{1}{2}
\left(\begin{array}{cc}
1+\cos\frac{5\pi}{4} & \sin\frac{5\pi}{4}  \\
\sin\frac{5\pi}{4} & 1-\cos\frac{5\pi}{4}
\end{array}\right),
\end{align}
The spectrum of $\rho_{AB}$ is $\{1/4,1/4,1/4,1/4,0,0,0,0\}$ (later we will see this spectrum remains invariant under arbitrary state rotations).
Moreover, note that the following relations hold
\begin{align}
\label{relation1}
\sin(\frac{\pi}{4})+\sin(\frac{3\pi}{4})+\sin(\frac{5\pi}{4})+\sin(\frac{7\pi}{4})=0,\nonumber\\
\cos(\frac{\pi}{4})+\cos(\frac{3\pi}{4})+\cos(\frac{5\pi}{4})+\cos(\frac{7\pi}{4})=0,
\end{align}
Thus the reduced state $\rho_B=\frac{1}{4}(\rho_{00}+\rho_{01}+\rho_{10}+\rho_{11})=\frac{1}{2}I$.
Since the measurement is performed on the qubit (subsystem B), we need to evaluate the reduced states
of subsystem A conditioned on the measurements.

To go through all possible one-qubit projective measurements, we adopt the projectors $\Pi_{\pm}=\frac{1}{2}(I\pm \vec{a}\cdot\vec{\sigma})$
with $|\vec{a}|^2=a_1^2+a_2^2+a_3^2=1$ and $\vec{\sigma}=(\sigma_1,\sigma_2,\sigma_3)$ the standard Pauli matrices. Accordingly,
the post-measurement states are (in terms of the computational basis $\{|00\rangle,|01\rangle,|10\rangle,|11\rangle\}$)
\begin{align}
\rho_A^+ &=\frac{1}{2}\mathbf{diag}\{P_{00}^+,P_{01}^+,P_{10}^+,P_{11}^+\},\nonumber\\
\rho_A^- &=\frac{1}{2}\mathbf{diag}\{P_{00}^-,P_{01}^-,P_{10}^-,P_{11}^-\},
\end{align}
where
\begin{align}
\label{spectrum1}
P_{00}^\pm &=\frac{1}{4}\left[2\pm2\big(a_1\sin\frac{\pi}{4}+a_3\cos\frac{\pi}{4}\big)\right],\nonumber\\
P_{01}^\pm &=\frac{1}{4}\left[2\pm2\big(a_1\sin\frac{7\pi}{4}+a_3\cos\frac{7\pi}{4}\big)\right],\nonumber\\
P_{10}^\pm &=\frac{1}{4}\left[2\pm2\big(a_1\sin\frac{3\pi}{4}+a_3\cos\frac{3\pi}{4}\big)\right],\nonumber\\
P_{11}^\pm &=\frac{1}{4}\left[2\pm2\big(a_1\sin\frac{5\pi}{4}+a_3\cos\frac{5\pi}{4}\big)\right],
\end{align}
Recalling the Eqs. (\ref{relation1}), the corresponding probabilities are given as
\begin{align}
p_+ &=\frac{1}{4}(P_{00}^++P_{01}^++P_{10}^++P_{11}^+)=\frac{1}{2},\nonumber\\
p_- &=\frac{1}{4}(P_{00}^-+P_{01}^-+P_{10}^-+P_{11}^-)=\frac{1}{2},
\end{align}
In addition, for this static case, we observe that $P_{00}^-=P_{11}^+$, $P_{01}^-=P_{10}^+$, $P_{10}^-=P_{01}^+$, and $P_{11}^-=P_{00}^+$.
Therefore, the quantum conditional entropy $S(\rho|\{\Pi_B^k\})=\sum_k p_kS(\rho_A^k)=S(\rho_A^+)=S(\rho_A^-)$. Then the
quantum discord (before optimization) can be obtained
\begin{align}
\label{rough}
\widetilde{\mathcal{D}}_B(\rho)&=S(\rho_B)-S(\rho)+\sum_k p_kS(\rho_A^k)\nonumber\\
&=S(\rho_A^+)-1=-\sum_i\frac{1}{2}P_i^+\log_2(\frac{1}{2}P_i^+)-1\nonumber\\
&=-\frac{1}{2}\sum_iP_i^+\log_2(P_i^+),
\end{align}
where the sum is over $i=00,01,10,11$.

So far, we arrive at the analytical expression of quantum discord without optimization, and
next step is to search through all the parameters involved to find out the minimum value.
Note that the spectrums of post-measurement states $\rho_A^\pm$ are independent of
$a_2$. Intuitively, it would be a good choice if we let $a_2=0$ \cite{Datta}. Actually,
our intuition is correct and the reason for this is as follows. We can define the set of parameters
\begin{align}
\left\{\begin{array}{ccc}
a_1=\eta\cos\theta,\\
a_2=\pm\sqrt{1-\eta^2},\\
a_3=\eta\sin\theta,
\end{array}\right.
\end{align}
with two variables $0\leq\eta\leq1$ and $\theta\in[0,\pi)$. It is easy to see that $\eta$ turns into
a global coefficient before the parentheses in Eqs. (\ref{spectrum1}) and the smaller the value of $\eta$,
the closer the spectrum of $\rho_A^\pm$ gets to $\{1/4,1/4,1/4,1/4\}$, which implies
$S(\rho_A^\pm)$ will gradually increase. Therefore, it is essential to set $\eta=1$ and then
we only need to perform the optimization over one variable $\theta$.
\begin{figure}[htbp]
\begin{center}
\includegraphics[width=0.4\textwidth ]{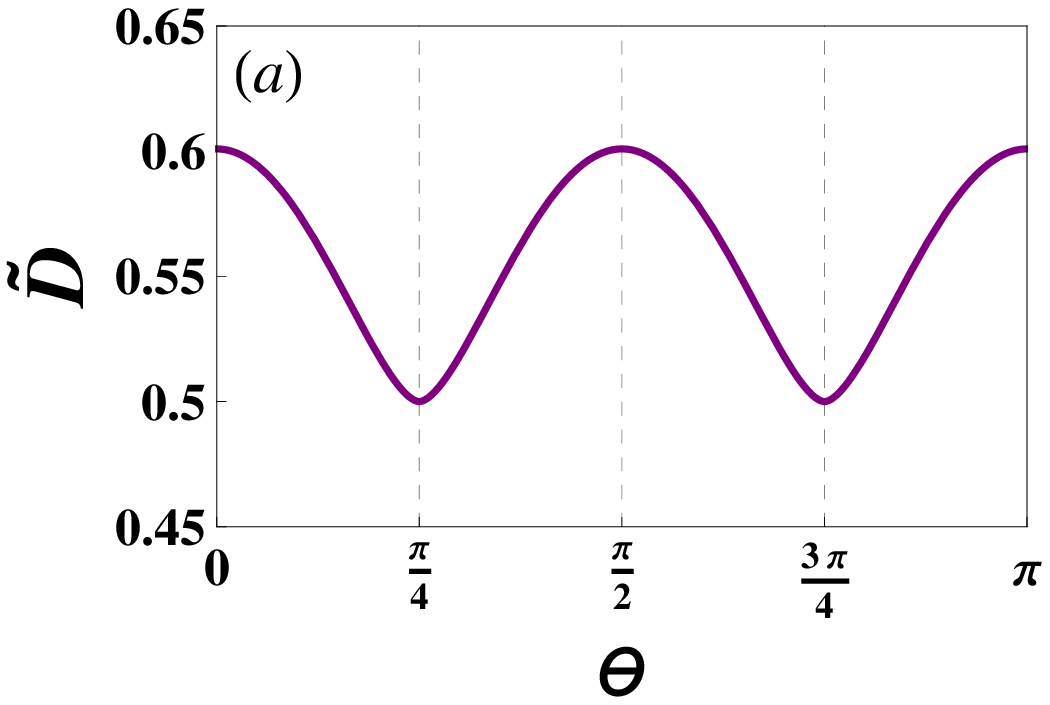}\\
\includegraphics[width=0.4\textwidth ]{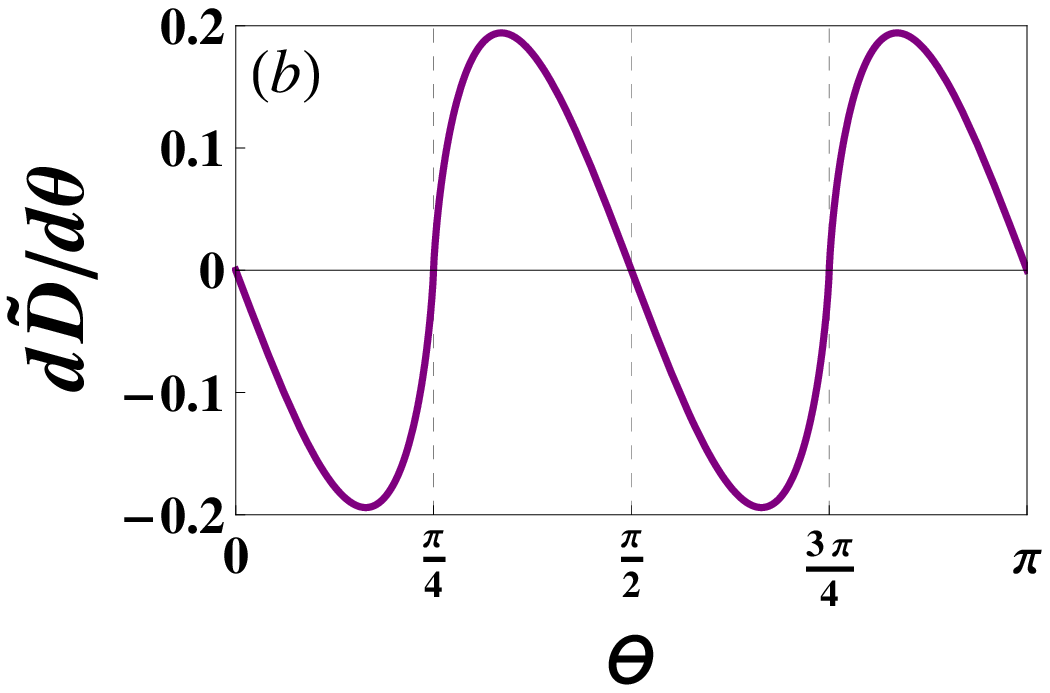}
\end{center}
\caption{(Color online) The $\widetilde{\mathcal{D}}_B(\rho)$ in Eq. (\ref{rough}) (a) and the first derivative of $\widetilde{\mathcal{D}}_B(\rho)$ (b)
as a function of the optimization parameter $\theta$.
}\label{QD1}
\end{figure}

Attempting to find the minimum value, we analyze the first derivative of $\widetilde{\mathcal{D}}_B(\rho)$,
which shows a periodic behavior, as plotted in Figure \ref{QD1}.
The data clearly shows that for $\theta=\frac{\pi}{4}$ or $\frac{3\pi}{4}$, $\widetilde{\mathcal{D}}_B(\rho)$ reaches the minimum
value $\frac{1}{2}$, which corresponds to the spectrum $\lambda(\rho_A^\pm)=\{1/2,1/4,1/4,0\}$. Now we know
that the exact value of quantum discord of the optimal $2\rightarrow1$ QRAC is
$\mathcal{D}_B(\rho)=\min_\theta\widetilde{\mathcal{D}}_B(\rho)=\frac{1}{2}$.

\subsection{Geometric discord}
In this subsection we try to assess the geometric discord. The key point is to represent state (\ref{state}) in Bloch form,
as done in Ref. \cite{Rana2012,Hassan2012}. First, let us briefly review the $SU(d)$ description of $d$-dimensional density operator.
The standard $SU(d)$ generators are natural extensions of the Pauli matrices (for qubits), which are also known as
the generalized Gell-Mann matrices (GGM) in higher-dimensional systems \cite{generator2}. They are defined as three different types of matrices and
for brevity we list these operators in the standard basis here
\begin{gather}
\mathcal{U}_{jk}=|j\rangle\langle k|+|k\rangle\langle j|,\nonumber\\
\mathcal{V}_{jk}=-i(|j\rangle\langle k|-|k\rangle\langle j|),\nonumber\\
\mathcal{W}_{l}=\sqrt{\frac{2}{l(l+1)}}\left(\sum_{j=1}^l|j\rangle\langle j|-l|l+1\rangle\langle l+1|\right),
\end{gather}
where $1\leq j<k\leq d$ and $1\leq l\leq d-1$.

To be consistent with the notations defined in Section II and also for the sake of simplicity,
we swap the subsystems A and B in Eq. (\ref{state}) and rephrase it as
\begin{align}
\label{swap-state}
\varrho_{AB}=&\frac{1}{4}(\rho_{00}\otimes|00\rangle\langle00|+\rho_{01}\otimes|01\rangle\langle01|\nonumber\\
&+\rho_{10}\otimes|10\rangle\langle10|+\rho_{11}\otimes|11\rangle\langle11|).
\end{align}
Note that the measurements are still performed on the qubit system (here means subsystem A) and this swap
procedure has no impact on the final results.

For subsystem A ($d=2$) the $SU(2)$ generators are the standard Pauli matrices, while for subsystem B ($d=4$)
the $SU(4)$ generators are a series of $4^2-1=15$ matrices \cite{generator2}. However, combining the Eqs. (\ref{decomposition})
(trace operator is involved) with the form of state (\ref{swap-state}), we immediately find that only three diagonal GGM
contribute to the calculation. In the standard basis they are given as
\begin{align}
\mathcal{W}_1=
\left(\begin{array}{cccc}
1 & 0 & 0 & 0 \\
0 & -1 & 0 & 0 \\
0 & 0 & 0 & 0 \\
0 & 0 & 0 & 0
\end{array}\right),\nonumber\\
\mathcal{W}_2=\frac{1}{\sqrt{3}}
\left(\begin{array}{cccc}
1 & 0 & 0 & 0 \\
0 & 1 & 0 & 0 \\
0 & 0 & -2 & 0 \\
0 & 0 & 0 & 0
\end{array}\right),\nonumber\\
\mathcal{W}_3=\frac{1}{\sqrt{6}}
\left(\begin{array}{cccc}
1 & 0 & 0 & 0 \\
0 & 1 & 0 & 0 \\
0 & 0 & 1 & 0 \\
0 & 0 & 0 & -3
\end{array}\right),
\end{align}
With these preparations, it is easy to obtain the Bloch vector $\vec{x}$
\begin{align}
\label{vector}
x_1&=\frac{1}{4}(\sin\frac{\pi}{4}+\sin\frac{3\pi}{4}+\sin\frac{5\pi}{4}+\sin\frac{7\pi}{4})=0,\nonumber\\
x_2&=0,\nonumber\\
x_3&=\frac{1}{4}(\cos\frac{\pi}{4}+\cos\frac{3\pi}{4}+\cos\frac{5\pi}{4}+\cos\frac{7\pi}{4})=0,
\end{align}
and the correlation matrix $T$
\begin{align}
T=
\left(\begin{matrix}
0 & 0 & \ldots & 0 & T_{11} & T_{12} & T_{13} \\
0 & 0 & \ldots & 0 & T_{21} & T_{22} & T_{23} \\
0 & 0 & \ldots & 0 & T_{31} & T_{32} & T_{33} \\
\end{matrix}\right),
\end{align}
where $T$ is a $3\times15$ matrix and later we will see that actually only 6 entries of
$T_{ij}=\frac{2\times4}{4}Tr(\varrho\sigma_i\otimes\mathcal{W}_j)$ can be nonzero
(in this static case there are five)
\begin{align}
\label{matrix}
T_{11}&=\frac{1}{2}(\sin\frac{\pi}{4}-\sin\frac{7\pi}{4}),\nonumber\\
T_{12}&=\frac{1}{2\sqrt{3}}(\sin\frac{\pi}{4}+\sin\frac{7\pi}{4}-2\sin\frac{3\pi}{4}),\nonumber\\
T_{13}&=\frac{1}{2\sqrt{6}}(\sin\frac{\pi}{4}+\sin\frac{7\pi}{4}+\sin\frac{3\pi}{4}-3\sin\frac{5\pi}{4}),\nonumber\\
T_{21}&=T_{22}=T_{23}=0,\\
T_{31}&=\frac{1}{2}(\cos\frac{\pi}{4}-\cos\frac{7\pi}{4})=0,\nonumber\\
T_{32}&=\frac{1}{2\sqrt{3}}(\cos\frac{\pi}{4}+\cos\frac{7\pi}{4}-2\cos\frac{3\pi}{4}),\nonumber\\
T_{33}&=\frac{1}{2\sqrt{6}}(\cos\frac{\pi}{4}+\cos\frac{7\pi}{4}+\cos\frac{3\pi}{4}-3\cos\frac{5\pi}{4}),\nonumber
\end{align}
Furthermore, we have
\begin{align}
G=\vec{x}\vec{x}^t+\frac{2TT^t}{n}=\frac{1}{2}
\left(\begin{array}{ccc}
1 & 0 & 0  \\
0 & 0 & 0 \\
0 & 0 & 1
\end{array}\right).
\end{align}
Thus the geometric discord of the optimal $2\rightarrow1$ QRAC is $\mathcal{D}_G=\frac{1}{16}$.

\section{DYNAMICAL BEHAVIOR OF QUANTUM DISCORD AND ITS CONNECTION WITH DIMENSION WITNESS}
In this section, we go a step further by demonstrating the dynamical behavior of quantum discord (and geometric discord)
concerning some possible state rotations. And more importantly, we illustrate the monotonic relationship between the
quantum discord and two-dimensional quantum witness \cite{Pawlowski2011,Li2011,Li2012}.

\subsection{Two-state case}
\begin{figure}[htbp]
\begin{center}
\includegraphics[width=.40\textwidth]{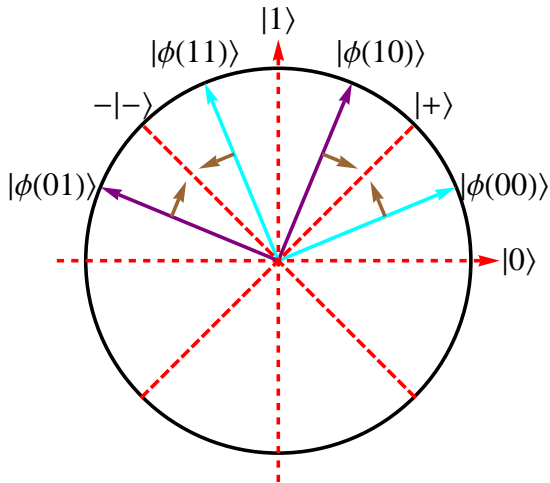} {}
\end{center}
\caption{(Color online) State rotations correspond to the two-state case: the orthogonal basis $\{|\phi(00)\rangle,|\phi(11)\rangle\}$
rotates as a unit counterclockwise with the angle $\delta$, while in contrast, another basis $\{|\phi(01)\rangle,|\phi(10)\rangle\}$
rotates as a unit clockwise with the same angle.
}
\label{two-state}
\end{figure}

To begin with, we first concentrate on the two-state case (see Figure \ref{two-state}). The state rotations
can be expressed as
\begin{align}
\label{transformation1}
\frac{\pi}{8}\rightarrow\frac{\pi}{8}+\delta,\quad \frac{7\pi}{8}\rightarrow\frac{7\pi}{8}-\delta,\nonumber\\
\frac{3\pi}{8}\rightarrow\frac{3\pi}{8}-\delta,\quad \frac{5\pi}{8}\rightarrow\frac{5\pi}{8}+\delta,
\end{align}
It is worth pointing out that since the double angle formula is applied in the above derivation of Eqs. (\ref{double}), in fact
the transformation can be viewed as $\frac{\pi}{4}\rightarrow\frac{\pi}{4}+2\delta, \frac{7\pi}{4}\rightarrow\frac{7\pi}{4}-2\delta,
\frac{3\pi}{4}\rightarrow\frac{3\pi}{4}-2\delta, \frac{5\pi}{4}\rightarrow\frac{5\pi}{4}+2\delta$.

Following the original definition of quantum discord, all we need is to evaluate the three terms $S(\rho_{AB})$, $S(\rho_{B})$
and $\sum_k p_kS(\rho_A^k)$ under this transformation. One can easily check that the spectrum of $\rho_{AB}$ remains unchanged
and in fact $\rho_{B}$ also keeps invariant because the following relations still hold
\begin{align}
\label{relation2}
\sum_{i=1}^4\sin(\phi_i)=\sum_{i=1}^4\cos(\phi_i)=0,
\end{align}
with $\phi_0=\frac{\pi}{4}+2\delta$, $\phi_1=\frac{7\pi}{4}-2\delta$, $\phi_2=\frac{3\pi}{4}-2\delta$, $\phi_3=\frac{5\pi}{4}+2\delta$.
For the same reason, we have $p_+=p_-=\frac{1}{2}$ again. Therefore, the formula (\ref{rough}) can still be employed in this case, and
of course the optimization can be preformed only over $\theta$. As for geometric discord, it turns out that the coefficients in
Eq. (\ref{vector}) and (\ref{matrix}) stay the same and we only need to take the angle transformation into consideration.
The geometric discord can be analytically obtained
\begin{align}
\label{GD}
\mathcal{D}_G&=\frac{1}{8}\left\{1-\max\big[\sin^2(\frac{\pi}{4}+2\delta),\cos^2(\frac{\pi}{4}+2\delta)\big]\right\}\nonumber\\
&=\frac{1}{16}(1-|\sin4\delta|).
\end{align}
Obviously, when $\delta=0$, $\mathcal{D}_G$ reduces to the static situation.

\begin{figure}[htbp]
\begin{center}
\includegraphics[width=0.4\textwidth ]{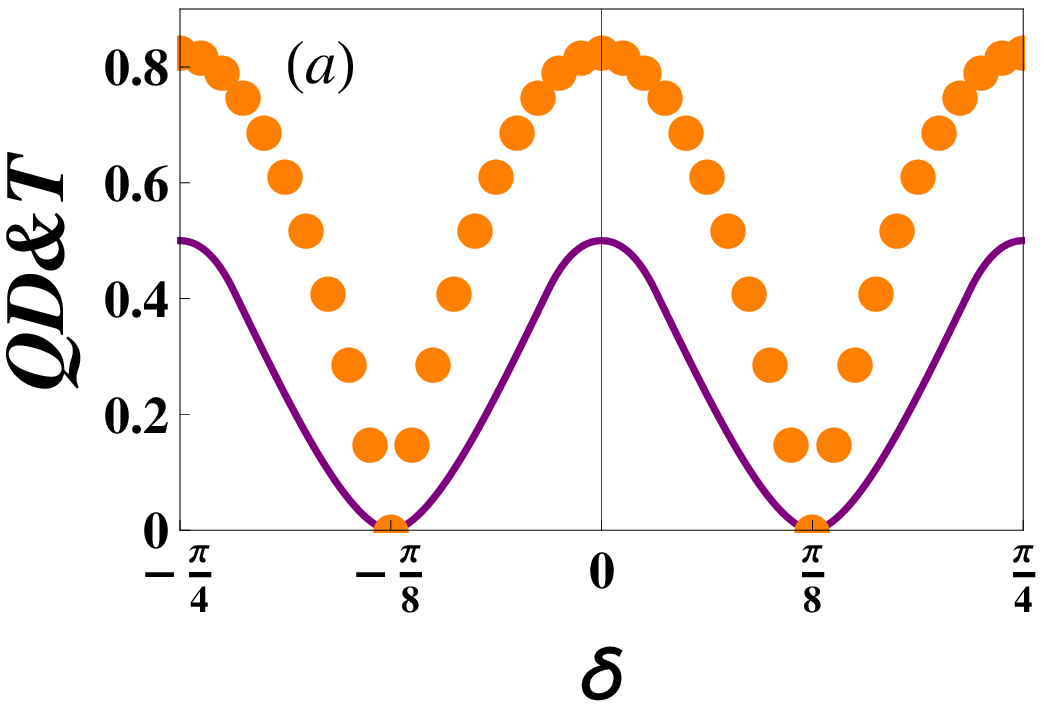}\\
\includegraphics[width=0.4\textwidth ]{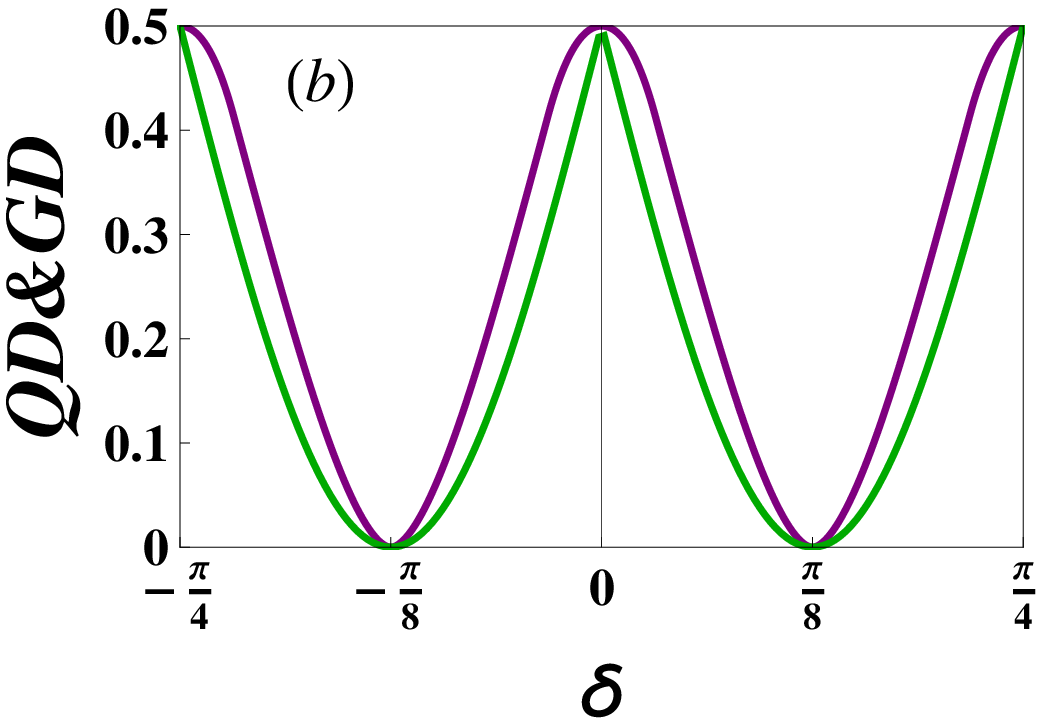}
\end{center}
\caption{(Color online) (a)The comparison between quantum discord (purple solid line) and two-dimensional quantum witness
(orange dotted line); (b)the comparison between quantum discord (purple solid line) and geometric discord (green solid line).
Note that here we actually plot $T-2$ \cite{T-2} and $8\mathcal{D}_G$ for clarity.
}\label{QD2}
\end{figure}

In our previous work \cite{Li2011,Li2012}, Li \textit{et al.} proposed a semi-device-independent random-number expansion
scenario with the help of $2\rightarrow1$ QRAC protocol, where the genuine randomness is certified by
the two-dimensional quantum witness violation $T$, which was first introduced by M. Paw{\l}owski and N. Brunner \cite{Pawlowski2011}
\begin{align}
\label{witness}
T\equiv&+E_{00,0}+E_{00,1}+E_{01,0}-E_{01,1}\nonumber\\
&-E_{10,0}+E_{10,1}-E_{11,0}-E_{11,1}\leq2\sqrt{2},
\end{align}
where $E_{a_1a_2,y}=P(b=0|a_1a_2,y)$ and $P(b|a,y)=Tr(\rho_a M_y^b)$ denotes the probability of Bob finding outcome $b$ when
he performed measurement $M_y$ and Alice prepared $\rho_a$.
The bound $T\leq2\sqrt{2}$ corresponds to the maximum violation of 2-dimensional witness
in the semi-device-independent black-box scenario, which is somewhat similar to the
maximal CHSH-inequality violation ($2\sqrt{2}$) by two-qubit states. Here, ``semi-device-independent'' indicates
only a two-dimensional system will be considered in this protocol and this bound $T\leq2\sqrt{2}$
was numerically presented in Ref. \cite{Li2011}. When $T=2\sqrt{2}$, it implies that Bob's success probability
of guessing any one bit of Alice's initial 2 bits is
\begin{align}
P_B=\frac{2+\sqrt{2}}{4}\approx0.85.
\end{align}
which is optimal in $2\rightarrow1$ QRAC.
For more details about dimension witnesses,
We would like to draw the reader's attention to the seminal paper by R. Gallego \textit{et al.} \cite{Gallego2010}.
Actually, Eq. (\ref{witness}) is a straightforward extension of the witness $I_3$ of Ref. \cite{Gallego2010}.

Within the $2\rightarrow1$ QRAC framework, we make two types of comparisons: (i) between QD and two-dimensional quantum witness
violation $(T-2)$ (Figure \ref{QD2} (a)); (ii) between QD and GD (Figure \ref{QD2} (b)). From these plots, there are several points
worth highlighting: (1) witness $T$ is monotonically related to QD, which implies the randomness guaranteed by
$T$ may have some connection with quantum correlations. However, we also notice that only when $T>2.64$ the
positive amount of randomness can be achieved \cite{Li2011}; (2) GD behaves highly monotonically with respect to QD as well, and
it indicates that in this case GD can also be viewed as a faithful measure of quantum correlation; (3) All the maximum or minimum values
of these three quantities occur simultaneously. For instance, when the rotation angle $\delta=\frac{\pi}{8}$, the orthogonal basis
$\{|\phi(00)\rangle,|\phi(11)\rangle\}$ and $\{|\phi(01)\rangle,|\phi(10)\rangle\}$ coincide at $\{|+\rangle,|-\rangle\}$. Accordingly,
the initial state $\rho_{AB}$ reduces to a \textit{classical-classical} state \cite{Piani2008}, which contains only classical correlations
and no quantum correlations; (4) Finally, we would like to point out that if one of the encoding states crosses the other due to the
rotations, the state ordering (encoding) must make corresponding changes. It is crucial for the calculation of $T$.

\subsection{Arbitrary rotations in $|0\rangle-|1\rangle$ plane}
In fact, we can numerically obtain dimension witness $T$ and quantum discord once the four rotation angles (corresponding to the four encoding states)
are \textit{specific}, applying the above algorithms raised in this paper. Analytical relationship can hardly be achieved and
this dilemma is mainly due to the definition and mathematical treatment of dimension witness $T$, since too many variables are involved in the optimization process. Actually, in some earlier papers \cite{Gallego2010,Brunner2008},
this difficulty has already been pointed out and they \textit{numerically} derived upper-bounds of dimension witnesses, using
\textit{semi-definite programming} (SDP) to solve the optimization problem.

We also note that one can obtain an analytic expression of geometric discord for the specific rotation discussed in subsection A.
Meanwhile, we realize that geometric discord may be more appropriate for characterizing quantum correlations
with respect to our topic, not only because of its mathematical simplicity but also due to its definition:
geometric discord is defined from the \textit{geometric} point of view. Therefore, we attempt to give the analytic formulation of geometric discord
for arbitrary rotations in real $|0\rangle-|1\rangle$ plane. The arbitrary rotation can be represented as
\begin{align}
\label{}
\frac{\pi}{8}\rightarrow\frac{\pi}{8}+\delta_1,\quad \frac{7\pi}{8}\rightarrow\frac{7\pi}{8}+\delta_2,\nonumber\\
\frac{3\pi}{8}\rightarrow\frac{3\pi}{8}+\delta_3,\quad \frac{5\pi}{8}\rightarrow\frac{5\pi}{8}+\delta_4,
\end{align}
This calculation is tedious and lengthy, but with the help of mathematical softwares,
we found that some terms can be eliminated and some terms can be collected. Finally we arrive at the following three
eigenvalues of $G=\vec{x}\vec{x}^t+\frac{2TT^t}{n}$
\begin{align}
\label{}
\lambda_1=&\frac{4-\sqrt{2\Delta}}{8},\nonumber\\
\lambda_2=&0,\nonumber\\
\lambda_3=&\frac{4+\sqrt{2\Delta}}{8},
\end{align}
where $\Delta=2+\cos4(\delta_1-\delta_4)+\cos4(\delta_2-\delta_3)-\cos4(\delta_1-\delta_2)
-\cos4(\delta_1-\delta_3)-\cos4(\delta_2-\delta_4)-\cos4(\delta_3-\delta_4)$.
It is easy to see that the example in subsection A can be rephrased as $\delta_1=\delta,\delta_2=-\delta,\delta_3=-\delta,\delta_4=\delta$.
Inserting these equations into the general formula, we can get
\begin{align}
\label{}
\lambda'_1=&\frac{1-|\sin4\delta|}{2},\nonumber\\
\lambda'_2=&0,\nonumber\\
\lambda'_3=&\frac{1+|\sin4\delta|}{2},
\end{align}
which exactly coincides with Eq. (\ref{GD}). Hence, the geometric discord for arbitrary rotations
can be obtained
\begin{align}
\mathcal{D}_G=\frac{1}{8}(1-\lambda_3)=\frac{1}{8}\lambda_1=\frac{4-\sqrt{2\Delta}}{64}.
\end{align}
It is remarkable that geometric discord reaches $\frac{1}{16}$ when $\delta_1=\delta_2=\delta_3=\delta_4=0$,
which is compatible with the optimal encoding strategy.
However, for arbitrary state encodings in Bloch sphere, we have performed numerical simulations in a more general situation of
the model and our results indicate that maximal geometric discord can not coincide with optimal $2\rightarrow1$ QRAC.
More details about numerical simulations are available in the Appendix.

\section{DISCUSSION AND CONCLUSION}
In this paper we have investigated quantum discord (and geometric discord) in quantum random access codes,
which have proved to be a valuable tool for a variety of applications in quantum information theory. We notice that
this model involves no entanglement at all and thus the usefulness of this protocol can not be attributed to entanglement.
However, our analysis highlights the presence of quantum discord in this protocol and it indicates that quantum discord
would be thought of as figure of merit for characterizing quantum nature in this communication model
since this model has no classical counterparts.

In two-state case, we explicitly elaborate the relations between
quantum discord and two-dimensional quantum witness following the method in \cite{Pawlowski2011,Li2011,Li2012}. Our
results show that the two quantities are monotonically related to each other and achieve the maximum or minimum values
under the same conditions. In addition, we also find that the geometric discord behaves highly monotonically
with respect to quantum discord in this case.

Furthermore, we derive an explicit analytical expression of the geometric discord if we restrict to state encodings in
real $|0\rangle-|1\rangle$ plane and it turns out that geometric discord reaches the maximal value for the optimal encoding strategy.
However, our numerical simulations reveal that for arbitrary state encodings in Bloch sphere,
maximal geometric discord could not coincide with optimal $2\rightarrow1$ QRAC, which is in sharp contrast
to the situation we encounter in $|0\rangle-|1\rangle$ plane. 
A clearer picture between quantum discord and dimension witnesses
remains an open question and deserves more investigation.

In view of these findings, we should note that there are many other interesting issues that remain to be addressed:
(i) it would be worthy of investigation in $3\rightarrow1$ QRAC, since Li \textit{et al.} pointed out that
the $3\rightarrow1$ QRAC is the most efficient semi-device-independent randomness-generation protocol known \cite{Li2012}.
(ii) although in two-state case witness $T$ is monotonically related to quantum discord, we notice that only
when $T>2.64$ the positive amount of randomness can be achieved \cite{Li2011}. Recently, experimental device-independent tests
of classical and quantum dimensions have been put forward \cite{Ahrens2012,Hendrych2012}. It is desirable to clarity the
intrinsic connection between dimension witnesses and quantum correlations. (iii) the relationship between quantum discord
and geometric discord may need further investigation, especially in high-dimensional systems. Since the validity of
geometric discord as a good measure for the quantumness of correlations has been questioned \cite{Piani2012}, its operational meaning
urgently need to be uncovered (a recent example has been reported by Streltsov \textit{et al.} \cite{Streltsov2012b}).

\begin{acknowledgments}
The author Y. Yao wishes to thank N. Brunner for his helpful comments and drawing our attention to
the seminal paper by R. Gallego \textit{et al.} \cite{Gallego2010}, and acknowledge the valuable suggestions of the anonymous referee.
This work was supported by the National Basic Research Program of China (Grants No. 2011CBA00200 and No. 2011CB921200),
National Natural Science Foundation of China (Grant NO. 60921091), and China Postdoctoral Science Foundation (Grant No. 20100480695).
\end{acknowledgments}
\appendix
\section{Numerical simulations in Bloch sphere}

Here we consider the geometric discord as a figure of merit. The key point to evaluate
geometric discord is to find the three eigenvalues $\lambda_1,\lambda_2,\lambda_3$ of the matrix
\begin{equation}
G=\vec{x}\vec{x}^t+\frac{2TT^t}{n},
\end{equation}
with $\lambda_1,\lambda_2,\lambda_3\in[0,1]$ and $TrG=\lambda_1+\lambda_2+\lambda_3=1$.
Indeed, for arbitrary encodings $|0\rangle-|1\rangle$ plane, we have analytically
obtained that
\begin{align}
\label{}
\lambda_1=&\frac{4-\sqrt{2\Delta}}{8},\nonumber\\
\lambda_2=&0,\nonumber\\
\lambda_3=&\frac{4+\sqrt{2\Delta}}{8},
\end{align}
where $\Delta=2+\cos4(\delta_1-\delta_4)+\cos4(\delta_2-\delta_3)-\cos4(\delta_1-\delta_2)
-\cos4(\delta_1-\delta_3)-\cos4(\delta_2-\delta_4)-\cos4(\delta_3-\delta_4)$.
It is worth noting that one of the eigenvalues is zero (which indicates that $\lambda_1\leq\frac{1}{2}$ and
$\lambda_3\geq\frac{1}{2}$) and the geometric discord
can be given as
\begin{align}
\mathcal{D}_G=\frac{1}{8}(1-\lambda_3)=\frac{1}{8}\lambda_1=\frac{4-\sqrt{2\Delta}}{64}\leq\frac{1}{16},
\end{align}
where the equality is satisfied if $\delta_1=\delta_2=\delta_3=\delta_4=0$, which just corresponds to the
optimal encodings. However, if we consider arbitrary encodings in Bloch sphere, the situation becomes
technically hard to solve and we can only turn to numerical simulations. Before proceeding, it is
intuitive to think that for arbitrary encodings probably the three eigenvalues are all greater than 0 since
more variables are involved in this case. If so, we can expect that
\begin{align}
\mathcal{D}_G=\frac{1}{8}\left(1-\max[\lambda_1,\lambda_2,\lambda_3]\right)
\leq\frac{1}{12}.
\end{align}
where the equality is satisfied if $\lambda_1=\lambda_2=\lambda_3=\frac{1}{3}$.

\begin{figure}[htbp]
\begin{center}
\includegraphics[width=.40\textwidth]{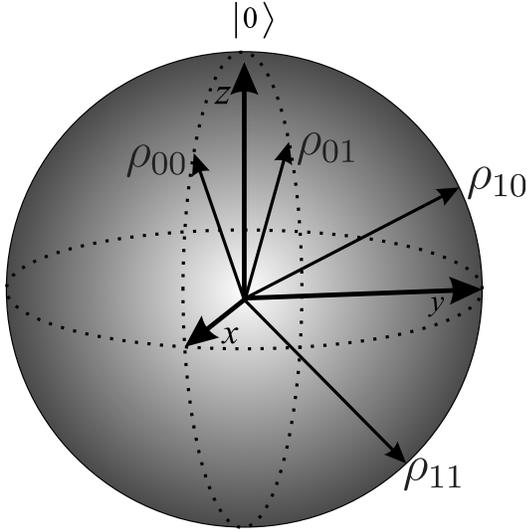} {}
\end{center}
\caption{(Color online) The Bloch sphere representation of arbitrary encodings.
}\label{bloch}
\end{figure}

In this situation, the Bloch sphere representation
can be employed as a very useful tool to run the simulations. A pure qubit
state can be represented as
\begin{align}
\label{pure}
|\psi\rangle=\cos\frac{\theta}{2}|0\rangle+e^{i\varphi}\sin\frac{\theta}{2}|1\rangle,
\end{align}
where $\theta\in[0,\pi]$ and $\varphi\in[0,2\pi]$. The Bloch vector for state (\ref{pure})
is $\vec{r}=(x,y,z)$, where the coordinates are given by
\begin{align}
\left\{\begin{array}{ccc}
x=\sin\theta\cos\varphi,\\
y=\sin\theta\sin\varphi,\\
z=\cos\theta,
\end{array}\right.
\end{align}
Since the phase factor $\varphi$ is involved in the representation, the four encoding states
may not lie in the same plane. However, up to local unitary equivalence, we can assume that
two of the states (for example, $\rho_{00}$ and $\rho_{01}$) are in the $|0\rangle-|1\rangle$ plane (or $x-z$ plane)
without loss of generality. Therefore, the four encoding states can be written as
\begin{align}
\label{encoding}
|\phi(00)\rangle &= \cos(\frac{\pi}{8}+\delta_1)|0\rangle+\sin(\frac{\pi}{8}+\delta_1)|1\rangle,\nonumber\\
|\phi(01)\rangle &= \cos(\frac{7\pi}{8}+\delta_2)|0\rangle+\sin(\frac{7\pi}{8}+\delta_2)|1\rangle,\nonumber\\
|\phi(10)\rangle &= \cos(\frac{3\pi}{8}+\delta_3)|0\rangle+e^{i\varphi_1}\sin(\frac{3\pi}{8}+\delta_3)|1\rangle,\nonumber\\
|\phi(11)\rangle &= \cos(\frac{5\pi}{8}+\delta_4)|0\rangle+e^{i\varphi_2}\sin(\frac{5\pi}{8}+\delta_4)|1\rangle,
\end{align}
where the six parameters $\delta_1,\delta_2,\delta_3,\delta_4\in[0,2\pi]$ and $\varphi_1,\varphi_2\in[0,2\pi]$
(see Figure \ref{bloch}).

After some algebra, we can obtain the Bloch vector $\vec{x}$
\begin{align}
x_1=&\frac{1}{4}\bigg[\sin(\frac{\pi}{4}+2\delta_1)+\sin(\frac{7\pi}{4}+2\delta_2)\nonumber\\
&+\sin(\frac{3\pi}{4}+2\delta_3)\cos\varphi_1+\sin(\frac{5\pi}{4}+2\delta_4)\cos\varphi_2\bigg],\nonumber\\
x_2=&\frac{1}{4}\bigg[\sin(\frac{3\pi}{4}+2\delta_3)\sin\varphi_1
+\sin(\frac{5\pi}{4}+2\delta_4)\sin\varphi_2\bigg],\nonumber\\
x_3=&\frac{1}{4}\bigg[\cos(\frac{\pi}{4}+2\delta_1)+\cos(\frac{7\pi}{4}+2\delta_2)\nonumber\\
&+\cos(\frac{3\pi}{4}+2\delta_3)+\cos(\frac{5\pi}{4}+2\delta_4)\bigg],
\end{align}
and the correlation matrix $T$
\begin{align}
T=\frac{1}{2}
\left(\begin{matrix}
 T_{11} & T_{12} & T_{13} \\
 T_{21} & T_{22} & T_{23} \\
 T_{31} & T_{32} & T_{33} \\
\end{matrix}\right),
\end{align}
where
\begin{widetext}
\begin{align}
\label{matrix}
T_{11}&=\sin(\frac{\pi}{4}+2\delta_1)-\sin(\frac{7\pi}{4}+2\delta_2),\nonumber\\
T_{12}&=\frac{1}{\sqrt{3}}\left[\sin(\frac{\pi}{4}+2\delta_1)+\sin(\frac{7\pi}{4}+2\delta_2)
-2\sin(\frac{3\pi}{4}+2\delta_3)\cos\varphi_1\right],\nonumber\\
T_{13}&=\frac{1}{\sqrt{6}}\left[\sin(\frac{\pi}{4}+2\delta_1)+\sin(\frac{7\pi}{4}+2\delta_2)
+\sin(\frac{3\pi}{4}+2\delta_3)\cos\varphi_1-3\sin(\frac{5\pi}{4}+2\delta_4)\cos\varphi_2\right],\nonumber\\
T_{21}&=0,\nonumber\\
T_{22}&=\frac{1}{\sqrt{3}}\left[-2\sin(\frac{3\pi}{4}+2\delta_3)\sin\varphi_1\right],\\
T_{23}&=\frac{1}{\sqrt{6}}\left[\sin(\frac{3\pi}{4}+2\delta_3)\sin\varphi_1-3\sin(\frac{5\pi}{4}+2\delta_4)\sin\varphi_2\right],\nonumber\\
T_{31}&=\cos(\frac{\pi}{4}+2\delta_1)-\cos(\frac{7\pi}{4}+2\delta_2),\nonumber\\
T_{32}&=\frac{1}{\sqrt{3}}\left[\cos(\frac{\pi}{4}+2\delta_1)+\cos(\frac{7\pi}{4}+2\delta_2)
-2\cos(\frac{3\pi}{4}+2\delta_3)\right],\nonumber\\
T_{33}&=\frac{1}{\sqrt{6}}\left[\cos(\frac{\pi}{4}+2\delta_1)+\cos(\frac{7\pi}{4}+2\delta_2)
+\cos(\frac{3\pi}{4}+2\delta_3)-3\cos(\frac{5\pi}{4}+2\delta_4)\right],\nonumber
\end{align}
\end{widetext}

First, we resort to mathematical softwares and find that no analytical expressions could be obtained with respect to
these six parameters. Then we turn to run simulations by computer programming. Note that we have to
go through all the allowed ranges of the six parameters and it is a six-layer loop program, which
means that the smaller the step size is, the greater the accuracy obtained but the more time
it takes to do the calculations. For example, we first choose the step size as $\frac{\pi}{10}$ and
$\frac{\pi}{20}$ (it takes half an hour and about two days respectively to run the simulation).
Numerical results (see Table \ref{numerical}) reveal that using smaller step size would cause the geometric discord
to get closer and closer to $\frac{2}{3}=0.66\ldots$ (here we consider the normalized geometric discord $8\mathcal{D}_G$),
as we expected. Note that these values are already larger than $\frac{1}{2}$. Furthermore, we have realized that
directly reducing the calculating step size is not a satisfactory strategy since it takes too much time to
run the simulation. For instance, if we choose the step size to be $\frac{\pi}{100}$, it will cost us about
$\frac{1}{2}\times10^6=5\times10^5$ hours. Therefore, it is more desirable to search around some specific
points which have been identified by the simulation, with fine-grained step size (e.g., $\pi\times10^{-4}$).
One set of parameters is listed in Table 1 and the accuracy of the estimate of $\mathcal{D}_G$ will gradually increase
if we repeat the process for several more iterations (our later runs demonstrated that $8\mathcal{D}_G$
indeed approaches $\frac{2}{3}$). As a comparison, we have also obtained the corresponding 2-dimensional
quantum witnesses $T$ with with the Levenberg-Marquardt algorithm (see Table \ref{numerical}). From Table \ref{numerical}, we can see that
for arbitrary state encodings dimension witness $T$ does not behave monotonically with respect to $\mathcal{D}_G$.
Besides, if we let $\varphi_1=\varphi_2=0$ (which means all the four encoding states
are in the $|0\rangle-|1\rangle$ plane), simulation results show that $8\mathcal{D}_G$ reaches the maximum value $\frac{1}{2}$
when $\delta_1=\delta_2=\delta_3=\delta_4=0$, which is compatible with the analytical formula.

\begin{widetext}
\begin{center}
\begin{table}[ht]
\caption{Numerical results for arbitrary encodings in Bloch sphere.}
\vskip0.1cm
\begin{tabular}{|c|c|c|c|c|c|c|c|c|} \hline
Step size        & $\delta_1$ & $\delta_2$ & $\delta_3$ & $\delta_4$ & $\varphi_1$ & $\varphi_2$ & $8\mathcal{D}_G$ & $T$      \\ \hline
$\frac{\pi}{10}$ & $1.40\pi$   & $1.90\pi$   & $0.30\pi$   & $0.70\pi$   & $0.60\pi$    & $0.40\pi$    & 0.6090        &   1.9519     \\ \hline
$\frac{\pi}{20}$ & $0.35\pi$   & $1.90\pi$   & $0.45\pi$   & $1.55\pi$   & $0.60\pi$    & $0.35\pi$    & 0.6431        &   2.2740     \\ \hline
$\pi\times10^{-4}$  & $0.2509\pi$   & $0.1980\pi$   & $0.3909\pi$   & $1.6089\pi$   & $0.6928\pi$    & $0.3079\pi$    &  0.6649  &   1.1658     \\ \hline
\end{tabular}\label{numerical}
\end{table}
\end{center}
\end{widetext}

Finally, we can draw the conclusion that: (i) if we restrict to state encodings in real $|0\rangle-|1\rangle$ plane geometric discord
reaches the maximal value ($\frac{1}{16}$) for the optimal encoding strategy; (ii) however, for arbitrary state encodings in Bloch sphere,
maximal geometric discord (approaching $\frac{1}{12}$) could not coincide with optimal $2\rightarrow1$ QRAC, which is in sharp contrast
to the situation we encounter in $|0\rangle-|1\rangle$ plane.



\begin{thebibliography}{99}
\bibitem{Ollivier2001} H. Ollivier and W. H. Zurek, Phys. Rev. Lett. \textbf{88}, 017901 (2001).
\bibitem{Henderson2001} L. Henderson and V. Vedral, J. Phys. A \textbf{34}, 6899 (2001).
\bibitem{Dakic2010} B. Daki\'{c}, V. Vedral, and \u{C}. Brukner, Phys. Rev. Lett. \textbf{105}, 190502 (2010).
\bibitem{Luo2010a} S. Luo and S. Fu, Phys. Rev. A \textbf{82}, 034302 (2010).
\bibitem{Luo2008a} S. Luo, Phys. Rev. A \textbf{77}, 022301 (2008).
\bibitem{Modi2010} K. Modi, T. Paterek, W. Son, V. Vedral, and M. Williamson, Phys. Rev. Lett. \textbf{104}, 080501 (2010).
\bibitem{Oppenheim2002} J. Oppenheim, M. Horodecki, P. Horodecki, and R. Horodecki, Phys. Rev. Lett. \textbf{89}, 180402 (2002).
\bibitem{Ferraro2010} A. Ferraro, L. Aolita, D. Cavalcanti, F. M. Cucchietti, and A. Ac\'{i}n, Phys. Rev. A \textbf{81}, 052318 (2010).
\bibitem{Datta2008} A. Datta, A. Shaji, and C. M. Caves, Phys. Rev. Lett. \textbf{100}, 050502 (2008).
\bibitem{Lanyon2008} B. P. Lanyon, M. Barbieri, M. P. Almeida, and A. G. White, Phys. Rev. Lett. \textbf{101}, 200501 (2008).
\bibitem{Dakic2012} B. Dakic, Y. O. Lipp, X. Ma, M. Ringbauer, S. Kropatschek, S. Barz, T. Paterek, V. Vedral,
A. Zeilinger, \u{C}. Brukner, and P. Walther, arXiv:1203.1629 (2012).
\bibitem{Tufarelli2012} T. Tufarelli, D. Girolami, R. Vasile, S. Bose, G. Adesso, arXiv:1205.0251 (2012).
\bibitem{Streltsov2012a} A. Streltsov, H. Kampermann, and D. Bru{\ss}, Phys. Rev. Lett. \textbf{108}, 250501 (2012).
\bibitem{Chuan2012} T. K. Chuan, J. Maillard, K. Modi, T. Paterek, M. Paternostro, and M. Piani, arXiv:1203.1268 (2012).
\bibitem{Bennett1984} C.H. Bennett and G. Brassard, in Proceedings of IEEE International Conference on Computers, Systems and Signal Processing,
Bangalore, India (IEEE, New York, 1984), pp. 175-179.
\bibitem{Wiesner1983} S. Wiesner, SIGACT News \textbf{15}, 78 (1983).
\bibitem{Ambainis1999} A. Ambainis, A. Nayak, A. Ta-Shma, U. Vazirani, in Proceedings of the 31st Annual ACM Symposium
on Theory of Computing (STOC'99), pp. 376-383, 1999.
\bibitem{Ambainis2002} A. Ambainis, A. Nayak, A. Ta-Shma, and U. Vazirani, J. ACM \textbf{49}, 496 (2002).
\bibitem{Hayashi2006} M. Hayashi, K. Iwama, H. Nishimura, R. Raymond, and S. Yamashita, New J. Phys. \textbf{8}, 129 (2006).
\bibitem{Klauck2001} H. Klauck, in Proceedings of the 42nd IEEE Symposium on Foundations of Computer Science (FOCS'01), pp. 288, 2001.
\bibitem{Hayashi2007} M. Hayashi, K. Iwama, H. Nishimura, R. Raymond, S. Yamashita, in Proceedings of the 24th International Symposium on Theoretical
Aspects of Computer Science (STACS'07), pp. 610-621, 2007.
\bibitem{Pawlowski2009} M. Paw{\l}owski, T. Paterek, D. Kaszlikowski, V. Scarani, A. Winter, and M. Zukowski, Nature \textbf{461}, 1101 (2009).
\bibitem{Pawlowski2011} M. Paw{\l}owski and N. Brunner, Phys. Rev. A \textbf{84}, 010302(R) (2011).
\bibitem{Li2011} H-W. Li, Z-Q. Yin, Y-C. Wu, X-B. Zou, S. Wang, W. Chen, G-C. Guo, and Z-F. Han, Phys. Rev. A \textbf{84}, 034301 (2011).
\bibitem{Li2012} H-W. Li, M. Paw{\l}owski, Z-Q. Yin, G-C. Guo, and Z-F. Han, Phys. Rev. A \textbf{85}, 052308 (2012).
\bibitem{Pironio2010} S. Pironio et al., Nature (London) \textbf{464}, 1021 (2010).
\bibitem{Gallego2010} R. Gallego, N. Brunner, C. Hadley, and A. Ac\'{i}n, Phys. Rev. Lett. \textbf{105}, 230501 (2010).
\bibitem{Brunner2008} N. Brunner, S. Pironio, A. Ac\'{i}n, N. Gisin, A. A. M\'{e}thot, and V. Scarani, Phys. Rev. Lett. \textbf{100}, 210503 (2008).
\bibitem{Ambainis2008} A. Ambainis, D. Leung, L. Mancinska, M. Ozols, arXiv:0810.2937 (2008).
\bibitem{Luo2008b} S. Luo, Phys. Rev. A 77, 042303 (2008).
\bibitem{Ali2010} M. Ali, A. R. P. Rau, and G. Alber, Phys. Rev. A \textbf{81}, 042105 (2010).
\bibitem{Girolami2011} D. Girolami and G. Adesso, Phys. Rev. A \textbf{83}, 052108 (2011).
\bibitem{Chen2011} Q. Chen, C. Zhang, S. Yu, X. X. Yi, and C. H. Oh, Phys. Rev. A \textbf{84}, 042313 (2011).
\bibitem{Chitambar2011} E. Chitambar, arXiv:1110.3057 (2011).
\bibitem{Rana2012} S. Rana and P. Parashar, Phys. Rev. A \textbf{85}, 024102 (2012).
\bibitem{Hassan2012} A. S. M. Hassan, B. Lari and P. S. Joag, Phys. Rev. A \textbf{85}, 024302 (2012).
\bibitem{generator1} F. T. Hioe and J. H. Eberly, Phys. Rev. Lett. \textbf{47}, 838 (1981);
J. Schlienz and G. Mahler, Phys. Rev. A \textbf{52}, 4396 (1995);
\bibitem{Vinjanampathy2012} S. Vinjanampathy and A. R. P. Rau, J. Phys. A \textbf{45}, 095303 (2012).
\bibitem{Datta} This is exact the way that the author treated this problem in Ref. \cite{Datta2008}, but without explanation.
\bibitem{generator2} R. A. Bertlmann and P. Krammer, J. Phys. A \textbf{41}, 235303 (2008).
\bibitem{T-2} $T=2$ is the classical and quantum boundary for the 2-dimension witness $T$ (see \cite{Pawlowski2011,Gallego2010}),
which means that when $T>2$ the underlying system can \textit{only} be a quantum state. Therefore,
as we expected, the violation of $T$ (mathematically denoted as $T-2$) can be viewed as a signature of
the existence of quantum correlations. This is a key point in our analysis. From figure 5, it is clear to see that
``$T-2$'' keeps monotonic with regard to quantum discord. When $\delta=\frac{\pi}{8}$, the initial state reduces to a
classical-classical state, which contains only classical correlations and no quantum correlations. Notice at this
moment that $T-2=0$ and $D=0$ simultaneously.
\bibitem{Piani2008} M. Piani, P. Horodecki, and R. Horodecki, Phys. Rev. Lett. \textbf{100}, 090502 (2008).
\bibitem{Ahrens2012} J. Ahrens, P. Badzi\c{a}g, A. Cabello, and M. Bourennane, Nature Physics, \textbf{8}, 592-595 (2012).
\bibitem{Hendrych2012} M. Hendrych, R. Gallego, M. Mi\v{c}uda, N. Brunner, A. Ac\'{i}n and J. P. Torres, Nature Physics, \textbf{8}, 588-591 (2012).
\bibitem{Piani2012} M. Piani, arXiv:1206.0231 (2012).
\bibitem{Streltsov2012b} A. Streltsov, S. M. Giampaolo, W. Roga, D. Bru{\ss}, F. Illuminati, arXiv:1206.4075 (2012).
\end{thebibliography}
\end{document}